# Icosahedral quasicrystal-enhanced nucleation in Al alloys fabricated by selective laser melting

C. Galera[1,2], M.L. Montero-Sistiaga[3], K. Vanmeensel[3], M. Godino-Martínez[4], J. Llorca[1,2,+], M.T. Pérez-Prado[1*+]

[1] Imdea Materials Institute, Calle Eric Kandel, 2, 28906 Getafe, Madrid, Spain.

[2] Department of Materials Science, Polytechnic University of Madrid/Universidad Politécnica de Madrid, E. T. S. de Ingenieros de Caminos, Madrid 28040, Spain.

[3] KU Leuven, Department of Materials Engineering, Kasteelpark Arenberg 44, Heverlee, B-3001, Belgium.

[4] ENGIE Laborelec, Rodestraat 125, Linkebeek, B1630, Belgium.

[*]Corresponding author; [+]Equal contribution

**Abstract**

Selective laser melting (SLM) is rapidly evolving to become a mainstream technology. However, the fundamental mechanisms of solidification and microstructure development inherent to the non-equilibrium conditions of this additive manufacturing method, which differ largely from those typical of conventional processing techniques, remain widely unknown. In this work, an in-depth characterization of the microstructure of Al7075 SLM processed samples, built from powder mixtures containing $ZrH_2$ microparticles, demonstrates the occurrence of icosahedral quasicrystal-enhanced nucleation during laser fabrication. This solidification mechanism, only observed to date in cast Al-Zn and yellow gold alloys containing minute additions of Cr (Kurtuldu et al., 2013) or Ti (Chen et al. 2018), and Ir (Kurtuldu et al., 2014), is evidenced by the presence of an abnormally high fraction of twin boundaries and of five-fold orientation symmetry between twinned nearest neighbors lying within a matrix of equiaxed, randomly textured, ultrafine grains. This research attests to the wide range of possibilities offered by additive manufacturing methods for the investigation of novel physical metallurgy phenomena as well as for the design of advanced metals.

**1. Introduction**

Additive manufacturing (AM), or 3D printing of metals [1-4], one of the key technologies powering the Industry 4.0 revolution, allows for a significant reduction in weight, freedom of design, component customization, cutback of lead time facilitated by in-house production, and decreased waste. As such, AM is shifting the basis for competition in the fabrication of medical implants and aeronautical components, among others, thus disrupting traditional supply chains [5].

AM involves several out-of-equilibrium steps along its process chain. The feedstock for AM is a fine metallic powder, produced by rapid solidification, mostly via gas atomization, a process that involves cooling rates of the order of $10^5$-$10^6$ K/s [6]. Ideally, powders for AM should be spherical and <u>limited in</u> oxides in order to ensure good flowability and permeability. Although the optimum powder sizes are technique specific, in general, narrow size distributions and small particle sizes are desirable [7]. AM fabrication by selective laser melting (SLM), a widely spread and very versatile technique, consists of melting and rapid solidification of such metal powders as the laser scans selected areas of the powder layer following a predefined CAD design [8]. Here, again, the cooling rates range between $10^5$-$10^7$ K/s. It seems, therefore, reasonable that the solidification mechanisms and the microstructures resulting from AM differ from those obtained following conventional thermomechanical methods such as casting and wrought processing.

Despite increasing adoption and industrial investment on AM [9], several major hurdles are still preventing the tremendous potential of this technology to be fully unlocked. At least two critical impediments are related to materials. First, many structural alloys that are endowed with excellent mechanical behavior when processed by traditional manufacturing methods are susceptible to significant solidification cracking during AM processing [1], rendering useless components. Two notable examples of high performance alloys with poor printability by SLM are the high strength Al alloy 7075 [10,11] and the high temperature nickel superalloy MAR M247 [12-15]. A second materials-related hurdle limiting the tremendous potential of additive manufacturing is the large knowledge gap regarding the microstructures resulting from the complex non-equilibrium processes involved in AM processing.

Indeed, only a small number of aluminum alloys are currently utilized as raw materials for additive manufacturing [1,10,11]. The most common are hardenable $Al_{10}SiMg$ and eutectic $AlSi_{12}$. The mechanical properties of AM components manufactured from these two alloys are comparable to those of as-cast or high-pressure die cast samples, but are clearly inferior to those exhibited by wrought components made from high strength Al alloys such as Al7075, which is endowed with a yield strength exceeding 500 MPa and a ductility of 3 to 9 % [16]. SLM of the latter is unfortunately difficult due to low weldability as well as high reflectivity and low viscosity (a disadvantage that is common to most conventional Al alloys). In particular, thermal contraction during AM processing results in extensive crack formation, which leads to brittleness. Furthermore, the evaporation during laser melting of low melting point alloying elements such as Zn, which are critical to form hardening phases, contributes as well to the resulting poor mechanical behavior.

Several strategies have been implemented to reduce defects and thus to improve the SLM printability of Al alloys. These include, on the one hand, extrinsic approaches consisting, for example, on post-processing by hot isostatic pressing (HIP) [17,18], as well as on tuning processing parameters such as the scan strategy [19] and the laser power [20]. On the other hand, intrinsic attempts to reduce crack susceptibility include, first, alloy composition tuning [21-30]. Notorious examples of this strategy are the addition of Zr and Sc to Al-Mg alloys, inducing the formation of coherent $Al_3(Sc,Zr)$ precipitates during solidification in SLM, such as in Scalmalloy, proprietary of the Airbus group [21-25], as well as alloying with Si [26-28]. Hot crack reduction by functionalization with microparticles [31-33] has also been implemented with significant success. In their pioneer work, Martin et al. [33] demonstrated that crack susceptibility could be remarkably reduced by mixing Al alloy powders with microparticles containing nucleants, which induced grain refinement during processing.

Due to the complex phenomena taking place when the laser interacts with the powder bed, the relationship between the SLM processing parameters, the composition of the powders, the solidification mechanisms, and the resulting microstructures has still not been established. Significant efforts are, in particular, needed to have a clear understanding of the effect of alloying additions and microparticles on the defect structure, the grain size, the texture, the grain boundary nature, and the precipitate distribution of Al alloys resulting from SLM processing. Filling this knowledge gap is critical to derive guidelines for the design of Al alloys with a high structural performance by AM.

This work aims to investigate the microstructure of a SLM processed 7075 alloy, functionalized with microparticles, following an approach mimicking that developed by Martin et al. [33]. Al7075 powder-microparticle mixtures with different microparticle content are considered, and an exhaustive characterization of the effect of microparticle addition on the crystallite size and orientation, the grain boundary nature and the precipitate distribution, bridging micro- and nanoscales, is carried out. This work evidences the occurrence of icosahedral quasicrystalline enhanced nucleation during processing of the Al7075 alloy and thus highlights the possibility to leverage this solidification mechanism to tailor the fraction of twin boundaries and the precipitate distribution, thereby opening up the spectrum of possibilities for microstructural design by selective laser melting.

**2. Material and methods**

Figure 1 illustrates the morphology and size of the raw materials utilized for this investigation. In particular, Fig. 1 (a) shows the gas atomized Al 7075 powders, whose composition, measured by energy-dispersive X-ray spectroscopy (EDX) in a scanning electron microscope (SEM) (Helios NanoLab 600i, FEI), is detailed in Table 1. These powders, which were supplied by TLS Technik, are endowed with the characteristic spherical morphology. The apparent and tap density values are 1.33 and 1.54 g/cm$^3$, respectively. Three attempts were made to measure the flowability using a standardized Hall flowmeter. The powders were only able to flow once and otherwise remained trapped in the funnel. The flowability thus measured amounted to 109 s. Fig. 1(b) depicts the ZrH$_2$ microparticles, provided by Hongwu International Group LTD, which have irregular, angular shapes. The composition of the microparticles is also summarized in Table 1. The particle size distributions corresponding to the powders and the microparticles are compared in Fig. 1(c). For the Al alloy powders, $d_{10}$=24.4 μm, $d_{50}$=39.0 μm, and $d_{90}$=62.8 μm; for the ZrH$_2$ microparticles, $d_{10}$=1.63 μm, $d_{50}$=3.77 μm, and $d_{90}$=8.28 μm. Fig. 1 (d) illustrates the X-ray diffraction pattern corresponding to the powders and the microparticles. This measurement was carried out in a Panalytical Empyrean diffractometer, with Cu-kα radiation.

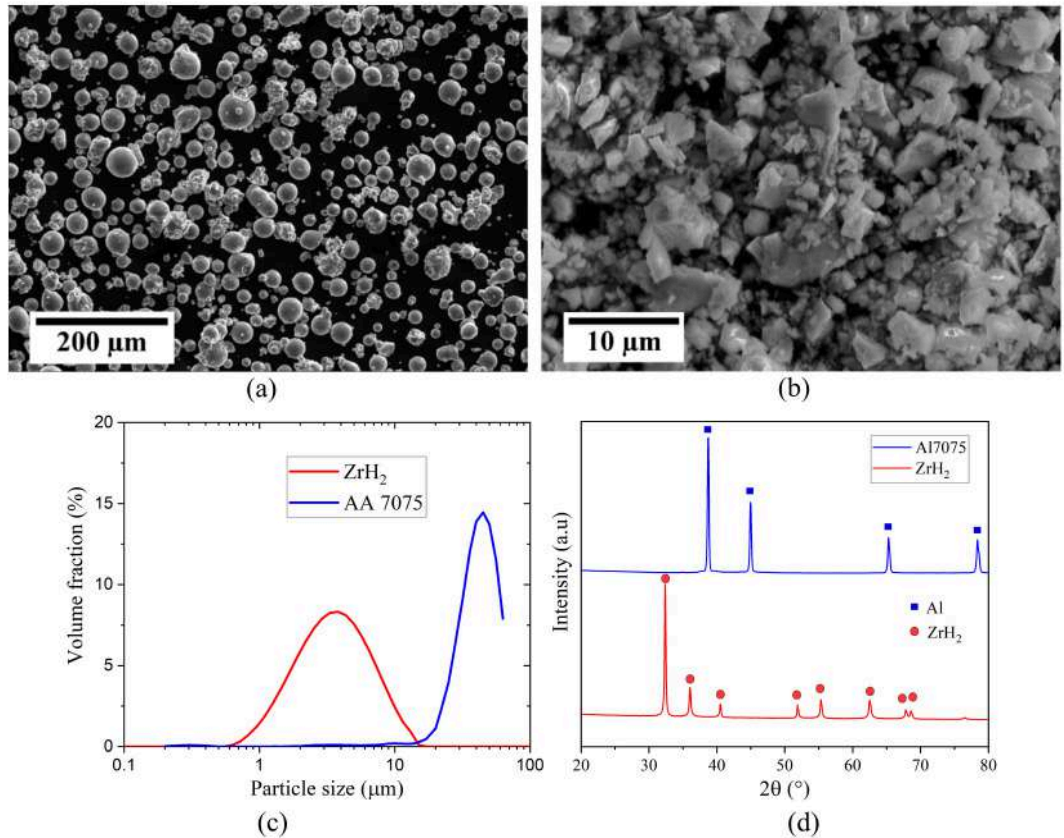

Figure 1. Secondary electron SEM micrographs illustrating the morphology of (a) the as-received Al 7075 powders and (b) the ZrH$_2$ particles; (c) comparison of the corresponding particle size distributions; (d) X-ray diffraction pattern of the Al7075 powders and the microparticles.

| (%wt)    | Al    | Zn   | Mg   | Fe   | Si   | Cr   | Cu   |
|----------|-------|------|------|------|------|------|------|
| *Average*  | 89.15 | 5.7  | 2.18 | 0.17 | 0.12 | 0.28 | 1.51 |
| **St. Dev.** | 1.41  | 0.96 | 0.18 | 0.15 | 0.22 | 0.07 | 0.28 |

| (%wt)    | Zr   | H   |
|----------|------|-----|
| *Average*  | 97.8 | 2.2 |
| **St. Dev.** | 1.1  | 1.1 |

Table 1. Composition of the Al 7075 powders and of the $ZrH_2$ microparticles.

Three Al 7075 and $ZrH_2$ mixtures containing, respectively, volume fractions of microparticles of 1, 2.5, and 5 %, were prepared using a Pulverisette 6 planetary mill without any grinding medium. Cycles consisting of 5 min of mixing at 200 rpm followed by a 10 min pause were repeated during 24 h for each mixture. Fig. 2 consists of several scanning electron microscopy (SEM) micrographs which illustrate the distribution of microparticles on the surface of the Al 7075 powders as a function of the microparticle content. As expected, the fraction of microparticles on the powder particles surfaces increases as the $ZrH_2$ volume fraction increases. In particular, it can be seen that while the mixture containing 1 vol. % of microparticles includes powder particles with microparticle-free surfaces, the surfaces of most powder particles are already covered with a homogeneous distribution of microparticles when the content increases to 2.5 vol. %. Fig. 2(d) shows a composition map obtained by energy-dispersive X-ray spectroscopy (EDX) in the SEM (Helios NanoLab 600i, FEI) in which the presence of the Zr-containing particles on the powder particle surfaces of the mix containing 2.5 vol. % $ZrH_2$ is evident.

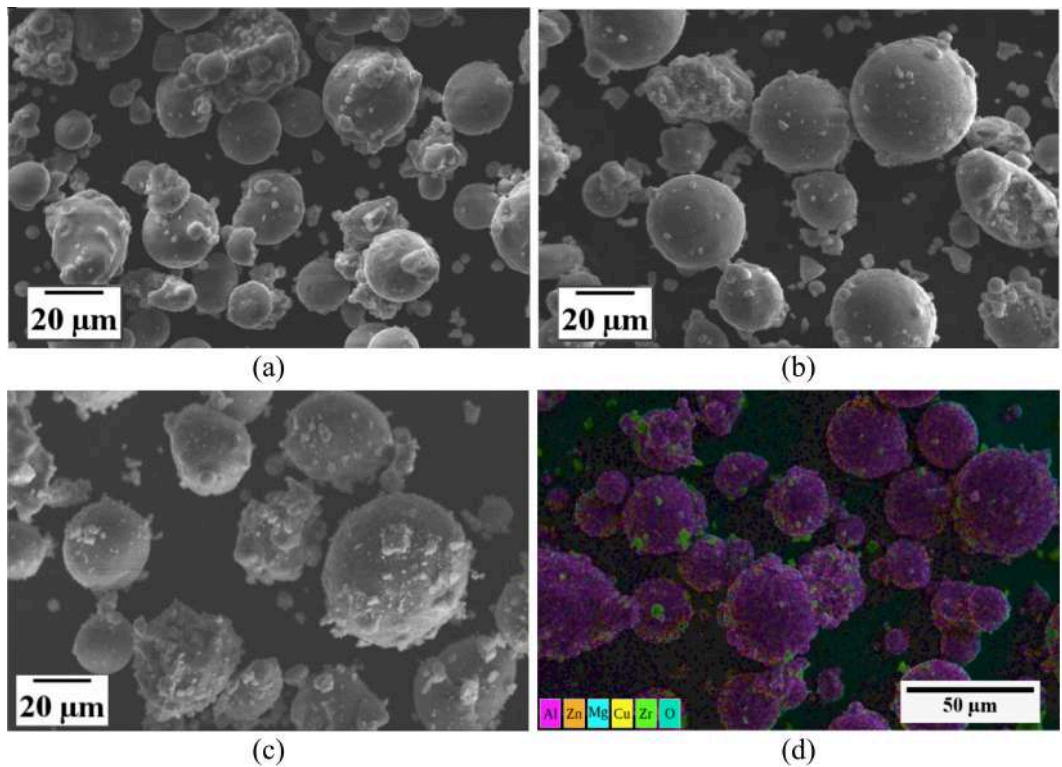

Figure 2. Distribution of microparticles on the surface of the Al 7075 powders in mixtures containing (a) 1, (b) 2.5, and (c) 5 vol. % ZrH$_2$; (d) EDX map of the mixture containing 2.5 vol. % ZrH$_2$.

SLM was carried out using as raw materials both the unmixed as-received Al 7075 alloy powders, as well as the powder-microparticle mixture containing 2.5 vol. % of ZrH$_2$, hereafter termed Al 7075+ZrH$_2$. The choice of 2.5 vol. % of microparticles was based on two reasons. On the one hand, it was anticipated that the homogeneous distribution of microparticles on the powder particles would contribute to reduce hot cracking by stimulating grain nucleation; on the other hand, in comparison with the mixture containing 5 vol. % of ZrH$_2$, it was expected that the reduced amount of hydrogen, that could potentially evaporate during processing, would lead to parts with higher density. Processing was performed in a Realizer 125 machine using a bidirectional scanning strategy, with a 67° rotation per layer. SLM parameter optimization was carried out by designing an experiment consisting of several combinations of processing parameters including the laser power (P), the hatch distance (H), and the scan speed (S), which are summarized in Table 2. The corresponding energy density (ED) values ranged from 32.9 to 123.7 J/mm$^3$. A square matrix of 0.5 x 0.5 x 0.3 cm$^3$ cubes was printed using the different processing conditions listed in Table 2. One cube was printed using each set of processing parameters. The densities of the Al 7075 and Al 7075+ZrH$_2$ cubes were measured following the Archimedes principle using water as a measuring medium and are also included in Table 2. The largest density values obtained range from 94 to 95%. Achieving higher densities was prevented by the entrapment of H$_2$ gas released from the ZrH$_2$ microparticles during processing. Optimum processing conditions were chosen as those rendering the highest density in the Al 7075+ZrH$_2$ cubes (94.7%). These processing conditions, highlighted using red bold characters in Table 2, include a laser power of 321 W, a hatch distance of 80 μm, and a scan speed of 1000 mm/s. The corresponding energy density is equal to 123.7 J/mm$^3$. The selected SLM parameter set gave rise to a 93.9 % density in the Al 7075 SLM processed sample.

The presence of defects in the printed parts was examined by optical microscopy (OM) in an Olympus BX-51 microscope. The microstructure and the microtexture of the Al 7075 and Al 7075+ZrH$_2$ samples manufactured using optimum SLM conditions were analyzed using a field emission gun scanning electron microscope (Helios NanoLab 600i, FEI). The microscope was equipped with an HKL electron backscattered diffraction (EBSD) system, a CCD camera, as well as both the Aztec and the Channel 5.0 data acquisition and analysis software packages. EBSD measurements were conducted using a step size of 0.2 μm, an accelerating voltage of 15 kV, and a beam current of 2.7 nA. The grain size was estimated from the EBSD maps by measuring the linear intercepts using grain boundaries with misorientation angles higher than 15º. Sample preparation for EBSD included several grinding steps with SiC paper followed by mechanical polishing with diamond pastes of increasingly finer particle sizes. Surface finishing was carried out by electropolishing in a Struers LectroPol-5 system using an electrolyte containing 30% vol. of nitric acid in methanol. Optimum electropolishing parameters differed for the Al 7075 and the Al 7075+ $ZrH_2$ printed samples. For the former, the electrolyte was kept at -15 °C and polishing was carried out at 15 V for 20 s; for the latter, the electrolyte was kept at -22 °C and polishing was carried out at 7 V for 20 s.

| Power (W) | Hatch (μm) | Speed (mm/s) | Energy density (J/mm³) | Density (%) Al 7075 | Density (%) Al 7075+ $ZrH_2$ | Power (W) | Hatch (μm) | Speed (mm/s) | Energy density (J/mm³) | Density (%) Al 7075 | Density (%) Al7075+ $ZrH_2$ |
|---|---|---|---|---|---|---|---|---|---|---|---|
| 222 | 60 | 1500 | 49,3 | 89,6 | 87,1 | 222 | 80 | 750 | 74,0 | - | 92,5 |
| 272 | 60 | 1500 | 60,4 | 89,5 | 88,9 | 272 | 80 | 750 | 90,7 | 94,2 | 93,3 |
| 321 | 60 | 1500 | 71,3 | 88,4 | 90 | 321 | 80 | 750 | 107,0 | 92,2 | 93,1 |
| 371 | 60 | 1500 | 82,4 | 90,2 | 90,3 | 371 | 80 | 750 | 123,7 | 92,5 | 93,3 |
| 222 | 70 | 1500 | 42,3 | 87,5 | 87,4 | 222 | 80 | 1000 | 55,5 | 92,5 | 91 |
| 272 | 70 | 1500 | 51,8 | 89,6 | 91,1 | 272 | 80 | 1000 | 68,0 | 92,6 | 94,2 |
| 321 | 70 | 1500 | 61,1 | 91,7 | 92,5 | *321* | *80* | *1000* | *80,3* | *93,9* | *94,7* |
| 371 | 70 | 1500 | 70,7 | 91,9 | 92,2 | 371 | 80 | 1000 | 92,8 | 94,5 | 92 |
| 222 | 80 | 1500 | 37,0 | 90,1 | 88,9 | 222 | 80 | 1250 | 44,4 | 92,7 | 90,7 |
| 272 | 80 | 1500 | 45,3 | 90,8 | 92,8 | 272 | 80 | 1250 | 54,4 | 92,2 | 93,8 |
| 321 | 80 | 1500 | 53,5 | 93,2 | 93,5 | 321 | 80 | 1250 | 64,2 | 93,7 | 92,2 |
| 371 | 80 | 1500 | 61,8 | 94,3 | 93,2 | 371 | 80 | 1250 | 74,2 | 93,2 | 93 |
| 222 | 90 | 1500 | 32,9 | 90,6 | 86,8 | | | | | | |
| 272 | 90 | 1500 | 40,3 | 91,7 | 89,7 | | | | | | |
| 321 | 90 | 1500 | 47,6 | 93,7 | 92,1 | | | | | | |
| 371 | 90 | 1500 | 55,0 | 94,1 | 92,9 | | | | | | |

Table 2. Selective laser melting processing parameters and the corresponding density of Al 7075 and Al 7075+$ZrH_2$ parts. In red, the combination of SLM parameters that yields the highest density values in the Al 7075+$ZrH_2$ parts.

The phase distribution of the processed samples was measured by X-ray diffraction (XRD) in a Philips X'pert-Pro Panalytical diffractometer furnished with a PW3050/60 goniometer and filtered Cu K$\alpha$ radiation. The surface area examined was about 1 cm$^2$. X-ray diffraction data were corrected for background and defocusing using the Philips X'pert software. Transmission electron microscopy (TEM) was utilized to examine the morphology and the composition of nanoprecipitates in the Al 7075 and Al 7075+ZrH$_2$ samples manufactured using optimum SLM parameters, as well as to detect their preferential nucleation sites. Bright field (BF) and Z-contrast imaging by high angle annular dark field scanning transmission electron microscopy (HAADF-STEM) were performed using a Talos F200X FEI TEM operating at 200 kV. TEM foils were prepared using a trenching-and-lift-out focused ion beam (FIB) technique with the aid of a Pt deposition system and an easy-lift needle. Rough milling of trenches around a 1.5 μm lamella was followed by thinning of the foil, using low beam currents, and by foil extraction and deposition onto a Cu TEM grid. The final step consisted in milling the foil to electron transparency (approximately 100 nm thickness).

Vickers hardness measurements were performed in the samples processed using the optimum SLM conditions in a HMV-6 Shimadzu microhardness indenter using a load of 0.2 kg and a dwell time of 10 s. A total number of 4 measurements were performed for each condition.

## 3. Results

*3.1 Micro, meso, and nanostructure of Al 7075 processed by SLM.*

The microstructure of the Al 7075 alloy, processed by SLM using the selected combination of processing parameters (P=321 W, H=80 μm, and S=1000 mm/s; density=93.9 %), was examined at several length scales. Table 3 illustrates the composition of the printed part. As expected, partial evaporation of the elements with low melting points, such as Mg and Zn, takes place during melting. Figure 3(a) illustrates the presence of a high density of defects, including long cracks parallel to the building direction (BD), as well as round pores resultant from gas entrapment, and irregular pores that are associated to insufficient melting. The presence of hot cracking in SLM processed Al 7075 has been widely reported earlier and our observations are thus consistent with previous works [10,11]. Fig. 3(b) depicts the X-ray diffractogram corresponding to the as-built alloy. It can be seen that, within the resolution of this technique, the only detectable phase is Al. However, the presence of a small volume fraction of second phases cannot be ruled out, as the detection limit of XRD is approximately 3 %.

| (%wt) | Al | Zn | Mg | Cu | Si | Cr | Fe |
|---|---|---|---|---|---|---|---|
| **Average** | 91.03 | 4.67 | 1.79 | 1.61 | 0.36 | 0.32 | 0.23 |
| **St. Dev.** | 0.77 | 0.74 | 0.07 | 0.09 | 0.08 | 0.05 | 0.07 |

Table 3. Composition of the as-built Al 7075 part.

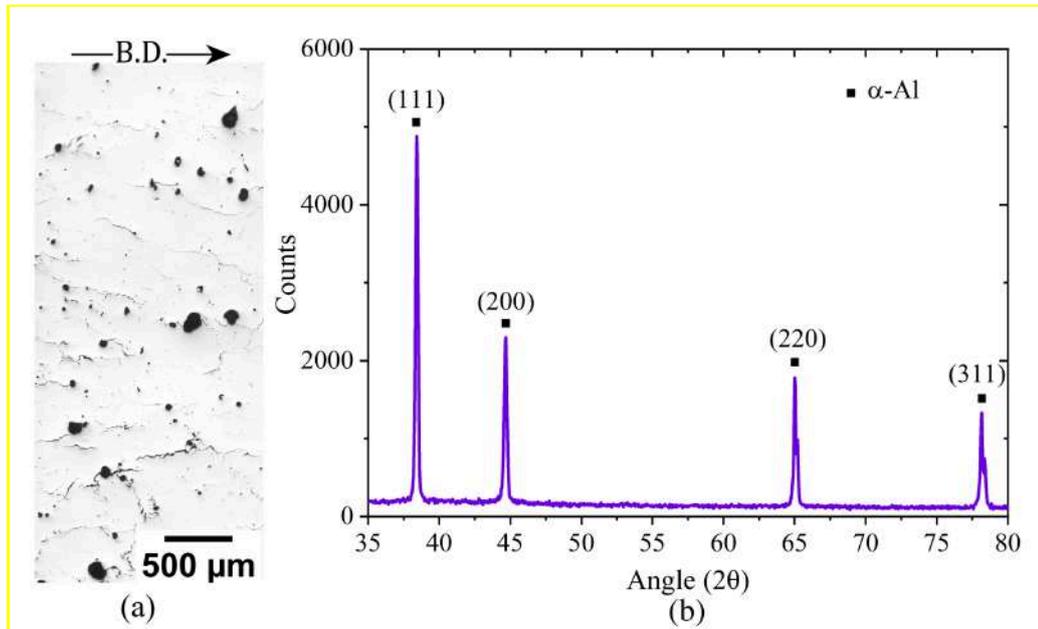

Figure 3. SLM processed Al 7075 alloy (as-built): (a) optical micrograph illustrating the presence of defects; (b) X-ray diffractogram showing the phase distribution.

Figure 4 illustrates the microtexture of the SLM processed Al 7075 alloy. Fig. 4(a) contains an EBSD inverse pole figure map in the BD in which boundaries with misorientation angles (θ) higher than 15º, hereafter termed high angle boundaries (HABs), have been colored in black, while boundaries with θ<15º, hereafter named low angle boundaries (LABs), are painted in white. This EBSD map shows that the microstructure is composed of grains with irregular shapes that are intermingled with small, equiaxed grains, and, second, that there are negligible variations of the microstructure along BD. Attempting to calculate the average grain size in such a heterogeneous microstructure would provide little information about its true nature. Fig. 5 therefore just aims to illustrate the distribution of linear intercepts measured from the EBSD map of Fig. 4(a) along the BD and a direction perpendicular to BD. Both distributions are rather wide, with intercepts ranging from less than 1 μm up to 100 μm along BD, and up to 50 μm along the perpendicular direction. Large grains exhibit significant intragranular lattice rotations, denoted by the presence of color gradients in their interiors, and contain a relatively high fraction of LABs. Additionally, Fig. 4(a) includes a series of inverse pole figures that illustrate the orientation of the BD in different areas of the specimen. These plots show that the texture of the printed part is overall very weak and that the <001> crystallographic direction exhibits only a slight alignment towards the BD. Fig. 4(b) compares the grain boundary misorientation histogram of the SLM processed Al 7075 alloy with the ideal MacKenzie distribution corresponding to a randomly oriented cubic polycrystal [34]. The two distributions show clear differences. Most importantly, the SLM processed material possesses an excess of LABs, which results from the need to accommodate the lattice distortions that originate as a consequence of the rapid cooling rates. In turn, the fraction of HABs in the processed sample is comparatively smaller than that of a cubic polycrystal with a similarly weak texture.

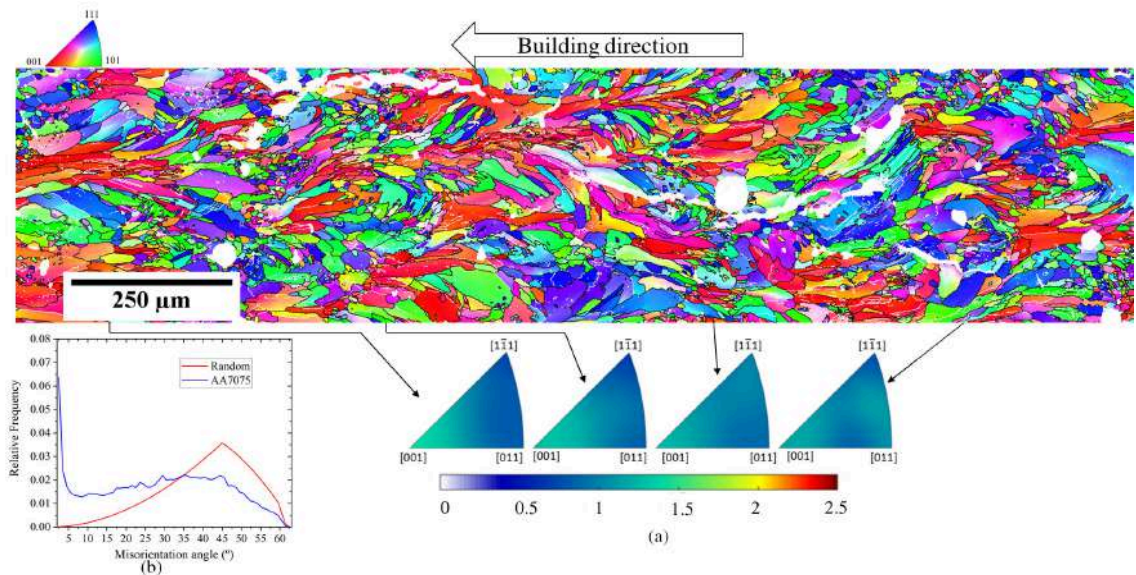

Figure 4. Microtexture of the SLM processed Al 7075 alloy (as-built). (a) EBSD inverse pole figure map in the building direction and inverse pole figures illustrating the orientation of the building direction with increasing build height. High angle boundaries ($\theta>15°$) have been colored in black, while low angle boundaries ($\theta<15°$) are painted in white; (b) comparison of the grain boundary misorientation distribution histogram of the SLM processed Al 7075 alloy with the ideal MacKenzie distribution corresponding to a randomly oriented cubic polycrystal [34].

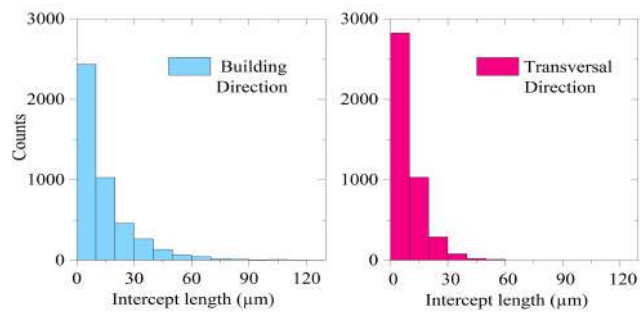

Figure 5. SLM processed Al 7075 alloy. Line intercepts along the BD and along a direction perpendicular to BD, measured from the EBSD inverse pole figure map of Fig. 4(a).

Fig. 6 is a TEM micrograph illustrating the presence of nanoprecipitates in the as-built Al 7075 alloy. These nanoprecipitates could not be unambiguously detected by conventional XRD due to their small size and comparatively low volume fraction. Precipitates are located both at grain boundaries, where they are endowed with elongated shapes, as well as at the grain interiors, where they have spherical morphologies. Fig. 7 consists of several HAADF- STEM element maps that show qualitatively the composition of the observed precipitates. It can be seen that while the grain boundaries are populated with Al-Cu intermetallics, the precipitates at the grain interiors are Al-Cu and, to a much smaller extent, of Al-Mg and Al-Mg-Cu particles. Fig. 8 illustrates the size distribution of Al-Cu particles, which range in size between 50 and 450 nm. Precipitation of $MgZn_2$ particles, which are the main strengthening phase in the Al 7075 alloy, was not observed.

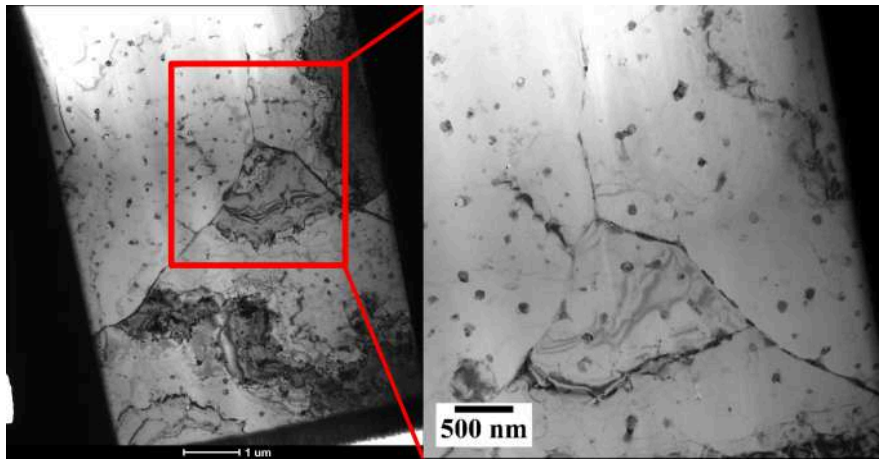

Figure 6. TEM micrograph illustrating the microstructure of the SLM processed Al 7075 alloy. (a) Image of the entire TEM lamella; (b) enlarged view of the area enclosed within the red rectangle in (a).

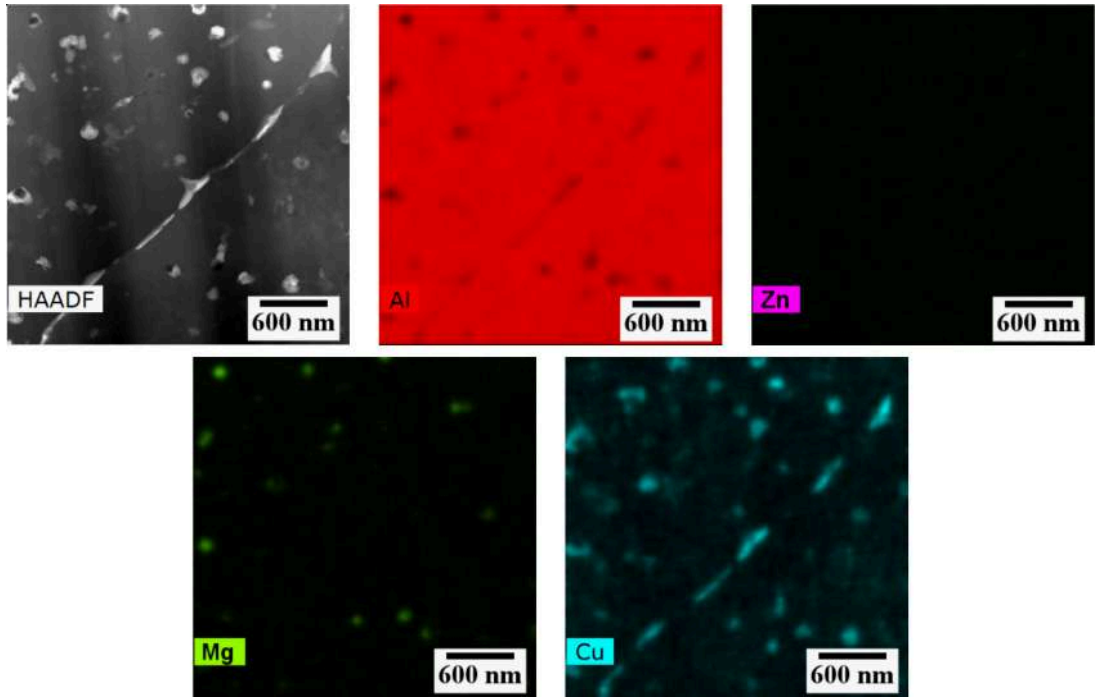

Figure 7. HAADF-STEM maps illustrating the composition of the precipitates that are present in the SLM processed Al 7075 alloy (as-built).

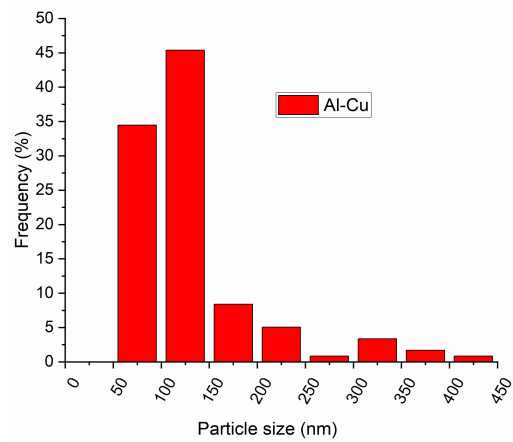

Figure 8. Size distribution of the precipitates present in the SLM processed Al 7075 alloy (as-built).

The average Vickers hardness of the SLM processed Al 7075 alloy amounted to 102.7±3.4 HV. This value is significantly lower than the hardness of the Al 7075 alloy in the T6 condition, which is approximately 175 HV [35], but is substantially higher than that corresponding to the same alloy in the solution treated condition (68 HV) [35].

*3.2 Micro, meso, and nanostructure of Al 7075+ZrH$_2$ processed by SLM.*

The microstructure of the Al 7075+ ZrH$_2$ mixture, processed by SLM using the selected combination of processing parameters (P=321 W, H=80 μm, and S=1000 mm/s; density=94.7 %), was examined at several length scales. Table 4 illustrates the composition of the printed part. As expected, partial evaporation of the elements with low melting points, such as Mg and Zn, takes place during melting. Fig. 9(a) is an optical micrograph that confirms the effectiveness of microparticle addition in preventing hot cracking, as observed earlier in [31]. The processed samples have, nevertheless, a rather high density of spherical pores due to hydrogen gas entrapment. Fig. 9(b) depicts the X-ray diffractogram corresponding to the built part. It can be seen that, within the resolution of this technique, the only detectable phase is Al. However, the presence of a small volume fraction of second phases again cannot be ruled out, as the detection limit of XRD is approximately 3 %.

| (%wt)    | Al    | Zn   | Zr   | Mg   | Cu   | Si   | Cr   | Fe   |
|----------|-------|------|------|------|------|------|------|------|
| **Average**  | 90.18 | 3.01 | 3.08 | 1.32 | 1.61 | 0.26 | 0.30 | 0.23 |
| **St. Dev.** | 0.51  | 0.39 | 0.31 | 0.09 | 0.12 | 0.04 | 0.04 | 0.05 |

Table 4. Composition of the SLM processed Al 7075+ZrH$_2$ part (as-built).

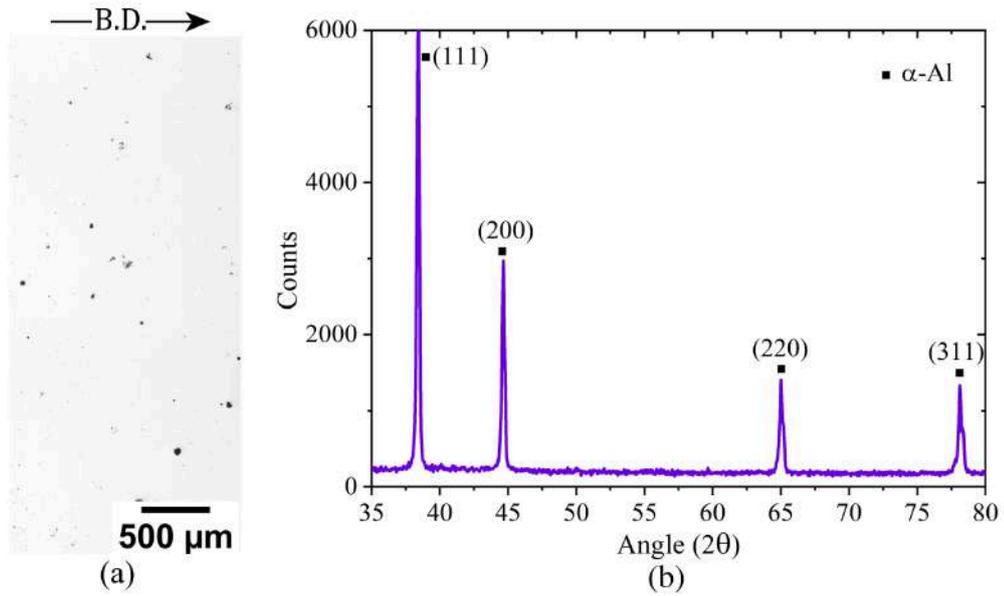

Figure 9. SLM processed Al 7075+ZrH$_2$ alloy (as-built): (a) optical micrograph illustrating the elimination of hot cracking; (b) X-ray diffractogram showing the phase distribution.

Fig.10 illustrates the microtexture of the SLM processed Al 7075+ZrH$_2$ alloy in the as-built condition. Figs. 10(a-c) are EBSD inverse pole figure maps in the BD which evidence, first, that the microstructure is mostly composed of fine, equiaxed grains, and, second, the homogeneity of the microstructure along BD, thus confirming that the effect of the addition of microparticles dominates over the effect of the increase in temperature with increasing build height. Fig. 11 illustrates the distribution of linear intercepts measured from the EBSD maps of Figs. 10(a-c) along BD and along a direction perpendicular to BD. The average intercept lengths are 0.76±0.46 μm, and 0.79±0.47 μm respectively. The dramatic grain refinement originated by the addition of the microparticles confirms the efficacy of the latter in enhancing grain nucleation. Figs. 10(a-c) include a series of inverse pole figures that illustrate the orientation of the BD at different heights of the specimen. These plots show that the texture of the printed part is overall basically random. Finally, Fig. 10(d) compares the grain boundary misorientation histogram corresponding to the top, middle, and bottom areas of the SLM processed Al 7075+ZrH$_2$ part with the ideal MacKenzie distribution corresponding to a randomly oriented cubic polycrystal [34]. As expected, the printed part has a larger fraction of LABs than the ideal MacKenzie distribution. Comparison of Figs. 4(b) and 10(d) reveals that the excess of LABs following SLM manufacturing becomes, however, less pronounced with the addition of microparticles, as grain subdivision is severely limited in grains with submicrometer sizes. Additionally, Fig. 10 (d) highlights the presence of an abnormally high number of twin boundaries ($\theta\sim60°$) in the SLM processed Al 7075+ZrH$_2$ part in comparison with that corresponding to other polycrystalline cubic materials with random textures, which are usually well described by the classical MacKenzie distribution [34]. These excess boundaries were not present in the microparticle-free SLM processed Al 7075 alloy, and are also not observed in Al alloys following other processing methods leading to a random texture, such as conventional casting [36,37,38], thermal treatments leading to recrystallization by particle stimulated nucleation [39,40], friction stir processing [41,42], or superplastic processing enabled by grain boundary sliding [41, 43]. The fraction of twin boundaries in microstructures resulting from such processes is less than 1 % [34,44], while that observed in the fine-grained Al 7075 + ZrH$_2$ microstructures fabricated by SLM is higher than 3%. This is surprising, since Al is well known to have a relatively high stacking fault energy, and thus a low propensity for twin boundary formation.

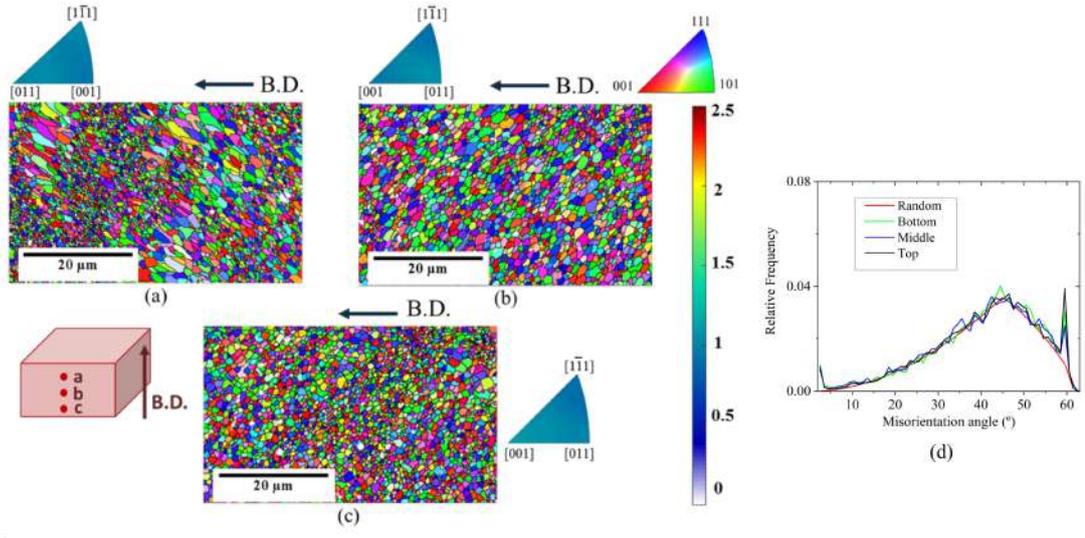

Figure 10. Microtexture of the as-built Al 7075+ZrH$_2$ alloy. (a-c) EBSD inverse pole figure maps in the building direction and inverse pole figures illustrating the orientation of the building direction with increasing build height. High angle boundaries (θ>15°) have been colored in black, while low angle boundaries (θ<15º) are painted in white; (d) Comparison of the grain boundary misorientation distribution histogram of the SLM processed Al 7075+ZrH$_2$ alloy with the ideal MacKenzie distribution corresponding to a randomly oriented face centered cubic polycrystal [34].

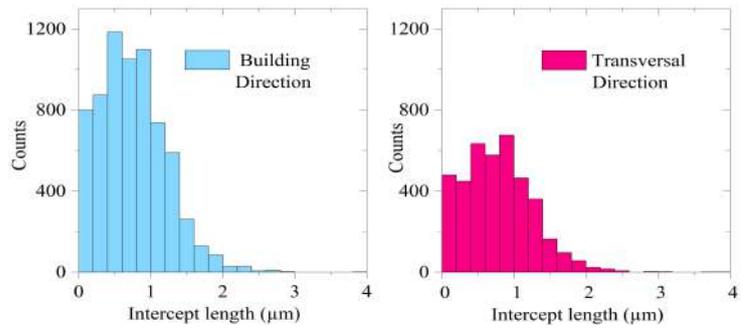

Figure 11. SLM processed Al 7075+ZrH$_2$ alloy. Line intercepts along the BD and along a direction perpendicular to BD, measured from the EBSD inverse pole figure maps of Figs. 10(a-c).

The influence of the SLM processing parameters on the porosity of the Al 7075+ZrH$_2$ alloy is illustrated in Fig. 12. This figure contains optical micrographs corresponding to samples processed with the following parameters: P=222 W, H=80 μm, and S=750 mm/s, ED=74 J/mm³ (Fig. 12(a)); P=321 W, H=80 μm, and S=1000 mm/s, ED=80.3 J/mm³, (Fig. 12(b)); P=222 W, H=80 μm, and S=1500 mm/s, ED=42.3 J/mm³, (Fig. 12(c)); P=321 W, H=80 μm, and S=750 mm/s, ED=107 J/mm³, (Fig. 12(d)); P=321 W, H=80 μm, and S=1500 mm/s, ED=53.5 J/mm³, (Fig. 12(e)). It can be seen that, irrespective of the processing conditions, hot cracking is fully eliminated. An analysis of the porosity distribution in these five samples, carried out by image, and including the size, shape, and density of pores, is presented in Table 5. It is shown that increasing the energy density is beneficial to reduce porosity.

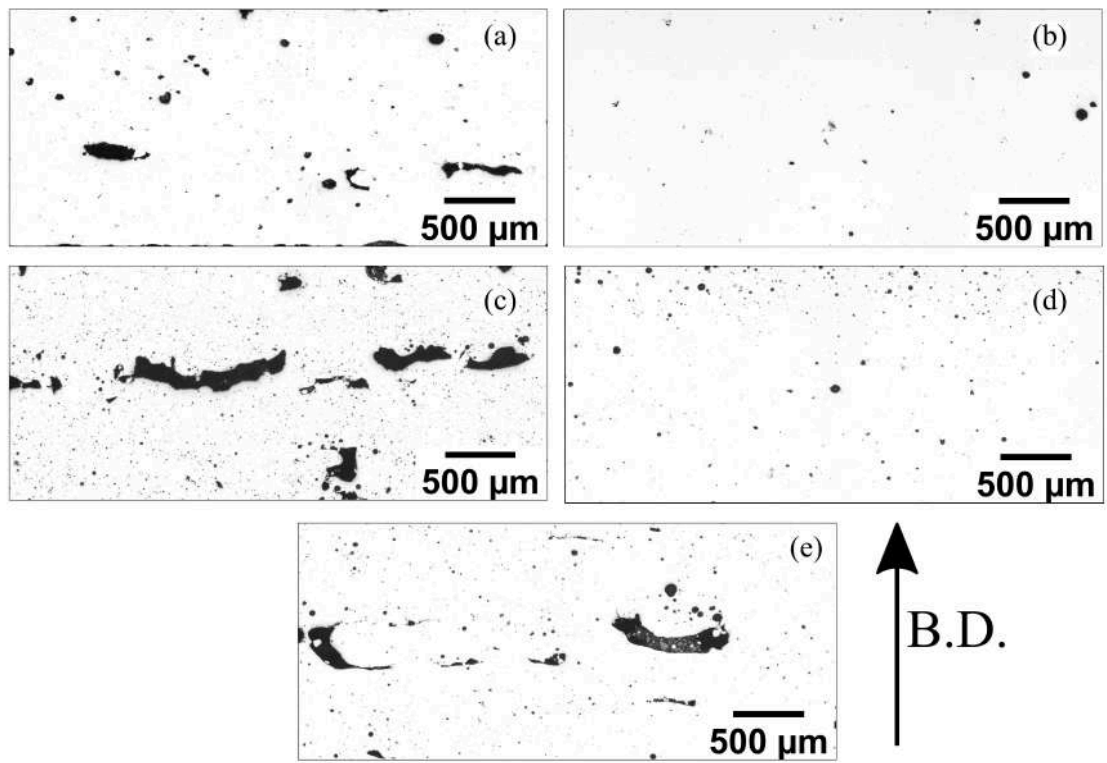

Figure 12. Influence of SLM processing parameters on the defect structure of the as-built Al 7075+ZrH$_2$ alloy. (a) P=222 W, H=80 μm, and S=750 mm/s, ED=74 J/mm³; (b) P=321 W, H=80 μm, and S=1000 mm/s, ED=80.3 J/mm³; (c) P=222 W, H=80 μm, and S=1500 mm/s, ED=42.3 J/mm³; (d) P=321 W, H=80 μm, and S=750 mm/s, ED=107 J/mm³; (e) P=321 W, H=80 μm, and S=1500 mm/s, ED=53.5 J/mm³.

| Sample | %Defects (Optical) | Average Fitted Ellipse Major Axis (μm) | Average Fitted Ellipse Minor Axis (μm) | Circularity | Number defects/ mm$^2$ | Volumetric Energy (J/mm$^3$) |
|---|---|---|---|---|---|---|
| a | **2.5** | 4.8 | 3.1 | 0.94 | 385 | 74 |
| b | **0.8** | 4.5 | 3.0 | 0.94 | 247 | 80 |
| c | **5.6** | 5.0 | 3.6 | 0.96 | 864 | 37 |
| d | **1.3** | 4.8 | 3.6 | 0.94 | 380 | 107 |
| e | **4.3** | 6.5 | 4.3 | 0.92 | 410 | 53 |

Table 5. Influence of SLM processing parameters on the porosity distribution in the as-built Al 7075+ZrH$_2$ alloy. (a) P=222 W, H=80 μm, and S=750 mm/s, ED=74 J/mm$^3$; (b) P=321 W, H=80 μm, and S=1000 mm/s, ED=80.3 J/mm$^3$; (c) P=222 W, H=80 μm, and S=1500 mm/s, ED=42.3 J/mm$^3$; (d) P=321 W, H=80 μm, and S=750 mm/s, ED=107 J/mm$^3$; (e) P=321 W, H=80 μm, and S=1500 mm/s, ED=53.5 J/mm$^3$.

The influence of the SLM processing parameters on the microstructural development of the Al 7075+ZrH$_2$ alloy is illustrated in Fig. 13. This figure includes EBSD inverse pole figure maps and inverse pole figures illustrating the orientation of the BD at the mid-layer of samples processed with the following parameters: P=222 W, H=80 μm, and S=750 mm/s, ED=74 J/mm$^3$ (Fig. 13(a)); P=321 W, H=80 μm, and S=1000 mm/s, ED=80.3 J/mm$^3$, (Fig. 13(b)); (c) P=222 W, H=80 μm, and S=1500 mm/s, ED=42.3 J/mm$^3$, (Fig. 13(c)); (d) P=321 W, H=80 μm, and S=750 mm/s, ED=107 J/mm$^3$, (Fig. 13(d)); (e) P=321 W, H=80 μm, and S=1500 mm/s, ED=53.5 J/mm$^3$, (Fig. 13(e)). Fig. 13(f) compares the grain boundary misorientation distribution histograms of the Al 7075+ZrH$_2$ alloy samples, processed using the five combinations of SLM parameters investigated, with the ideal MacKenzie distribution corresponding to a randomly oriented cubic polycrystal. Several observations can be inferred from this figure. First, irrespective of the processing conditions, significant grain refinement to submicrometer levels accompanied by texture randomization takes place during SLM due to the addition of the microparticles. Table 6 summarizes the average linear intercepts corresponding to the samples manufactured under all the SLM processing conditions investigated. Second, all SLM manufactured samples exhibit an abnormally high number of twin boundaries with respect to the MacKenzie distribution. Fig. 14 illustrates five boundary maps, corresponding to the five combinations of SLM parameters investigated, in which the excess twin boundaries have been plotted in red, and the remaining random HABs are painted in black. It can be clearly seen that the excess twin boundaries are not associated with twin lamellae grown in the interior of pre-existing grains, but that they mostly constitute interfaces between grains formed during the solidification process.

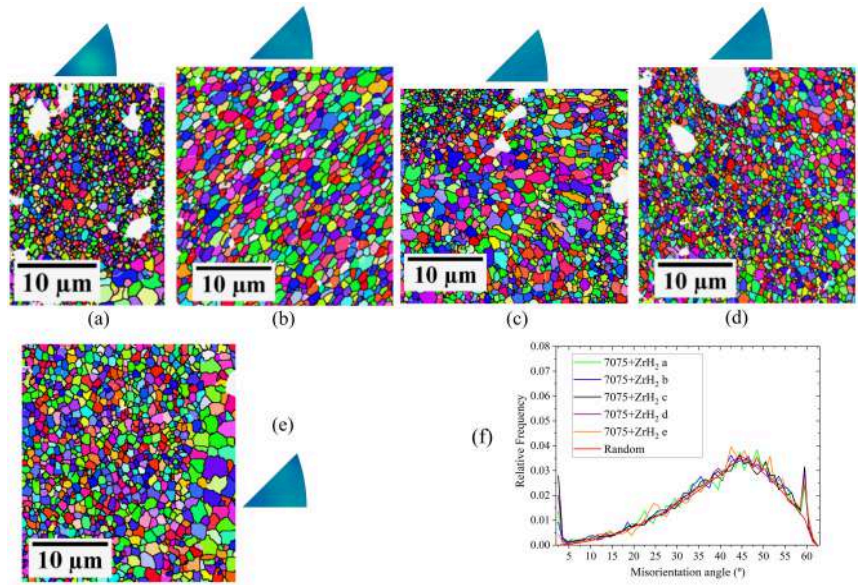

Figure 13. Influence of SLM processing parameters on the microtexture of the as-built Al 7075+ZrH$_2$ alloy. (a-e) EBSD IPF maps and inverse pole figures illustrating the orientation of the BD at the mid-layer of samples processed with the following parameters: (a) P=222 W, H=80 μm, and S=750 mm/s, ED=74 J/mm$^3$; (b) P=321 W, H=80 μm, and S=1000 mm/s, ED=80.3 J/mm$^3$; (c) P=222 W, H=80 μm, and S=1500 mm/s, ED=42.3 J/mm$^3$; (d) P=321 W, H=80 μm, and S=750 mm/s, ED=107 J/mm$^3$; (e) P=321 W, H=80 μm, and S=1500 mm/s, ED=53.5 J/mm$^3$. High angle boundaries (θ>15°) have been colored in black, while low angle boundaries (θ<15º) are painted in white; (f) comparison of the grain boundary misorientation distribution histograms of the Al 7075+ZrH$_2$ alloy with the ideal MacKenzie distribution [34].

| Average linear intercept | a | b | c | d | e |
|---|---|---|---|---|---|
| Building direction (μm) | 0,46 ±0,35 | 0,76 ±0,46 | 0,54 ±0,42 | 0,46 ±0,35 | 0,66 ±0,43 |
| Transversal direction (μm) | 0,50± 0,39 | 0,79 ±0,47 | 0,57 ±0,42 | 0,53 ±0,44 | 0,74 ±0,49 |

Table 6. Average linear intercepts along the BD and a long a direction perpendicular to BD at the mid-layer of as-built Al 7075+ZrH$_2$ samples processed with the following parameters: (a) P=222 W, H=80 μm, and S=750 mm/s; (b) P=321 W, H=80 μm, and S=1000 mm/s; (c) P=222 W, H=80 μm, and S=1500 mm/s; (d) P=321 W, H=80 μm, and S=750 mm/s; (e) P=321 W, H=80 μm, and S=1500 mm/s.

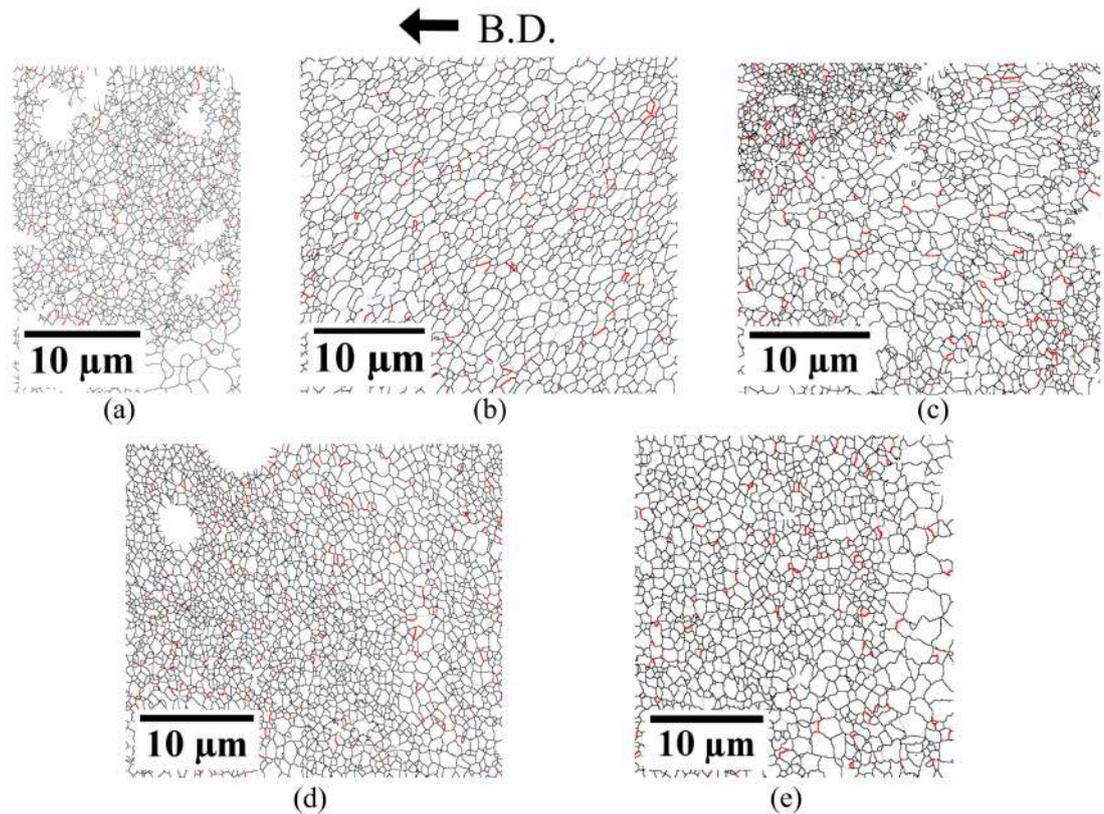

Figure 14. Twin boundaries developed during SLM processing of the Al 7075+ZrH$_2$ alloy under different processing conditions: (a) P=222 W, H=80 μm, and S=750 mm/s, ED=74 J/mm$^3$; (b) P=321 W, H=80 μm, and S=1000 mm/s, ED=80.3 J/mm$^3$; (c) P=222 W, H=80 μm, and S=1500 mm/s, ED=42.3 J/mm$^3$; (d) P=321 W, H=80 μm, and S=750 mm/s, ED=107 J/mm$^3$; (e) P=321 W, H=80 μm, and S=1500 mm/s, ED=53.5 J/mm$^3$. Twin boundaries (θ~60°) have been colored in red, while all other random high angle boundaries (θ>15º) are painted in black.

Figures 15-17 illustrate the precipitate distribution of the SLM processed Al 7075+ZrH$_2$ alloy. TEM revealed the presence of a wide range of precipitates, which could not be detected by XRD. First, it can be seen in Fig. 15 that grain boundaries are populated both with elongated precipitates as well as with clusters of round particles (highlighted using blue arrows); in turn, in grain interiors some square-shaped precipitates (highlighted with yellow arrows) and also some round particles are present.

Additionally, grain interiors are populated with a dispersion of very small nanoprecipitates (Fig. 17). Figs. 16 and 17 consist of several HAADF- STEM element maps that show qualitatively that the grain boundary precipitates and the round precipitates at grain interiors are mostly Al-Cu particles (Fig. 16) while the square-shaped precipitates at the grain interiors are Al-Zr intermetallics (Fig. 16). Finally, the dispersion of nanoprecipitates at the grain interiors is formed by both Al-Cu and Al-Zr particles (Fig. 17). Fig. 18 illustrates the size distribution of the different particles present in the SLM processed Al 7075+$ZrH_2$ alloy. Precipitation of $MgZn_2$ particles, which are the main strengthening phase in the Al 7075 alloy, was rarely observed.

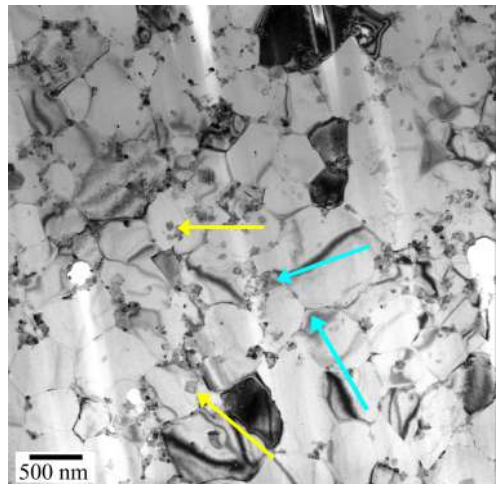

Figure 15. TEM micrograph illustrating the microstructure of the SLM processed Al 7075+$ZrH_2$ alloy. The yellow arrows point to square-shaped precipitates; the blue arrows point to more irregularly shaped particles, usually located at grain boundaries.

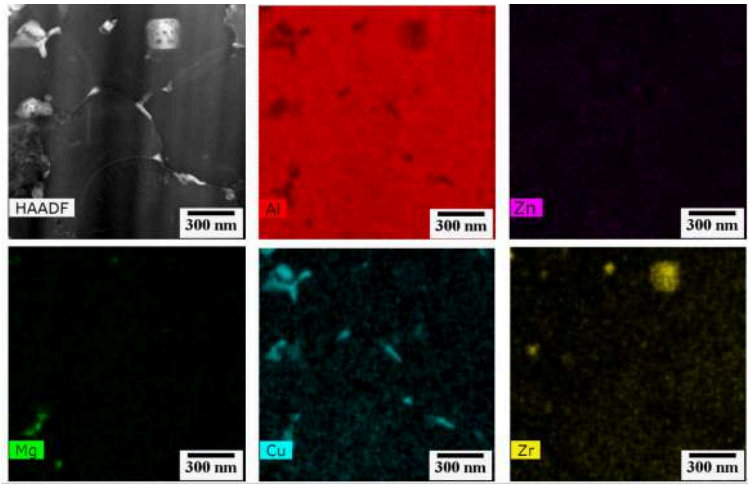

Figure 16. HAADF-STEM maps illustrating the precipitate composition in the as-built Al 7075+ZrH$_2$ alloy.

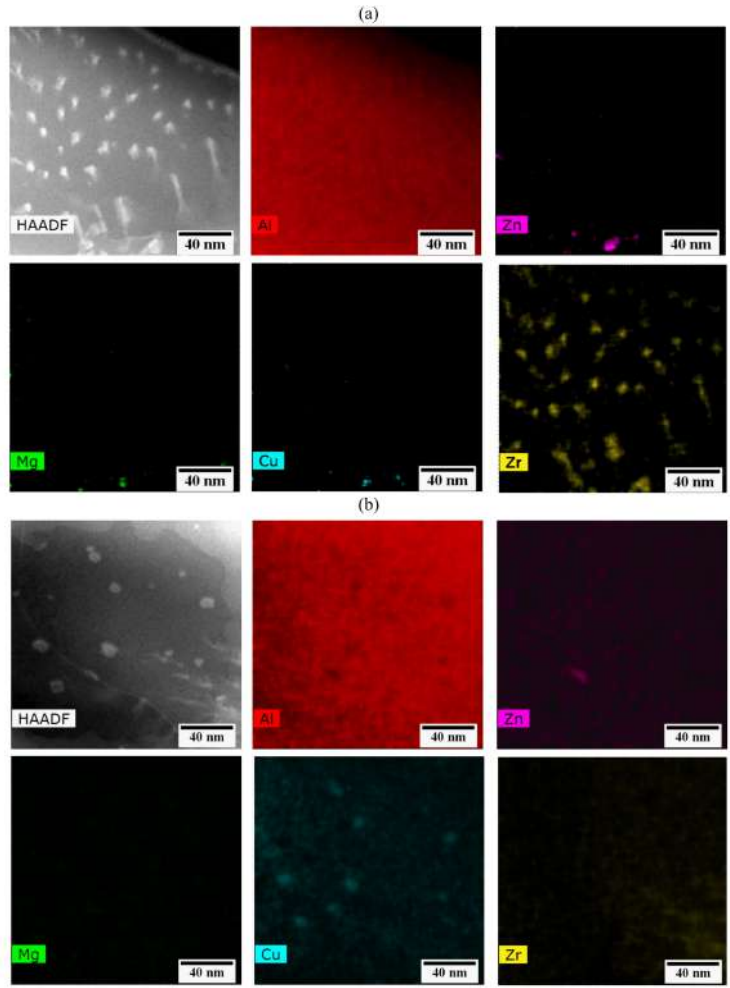

Figure 17. HAADF-STEM maps illustrating the composition of the (a) Al-Zr and (b) Al-Cu nanoprecipitates that are present at the grain interiors in the as-built Al 7075+ZrH$_2$ alloy.

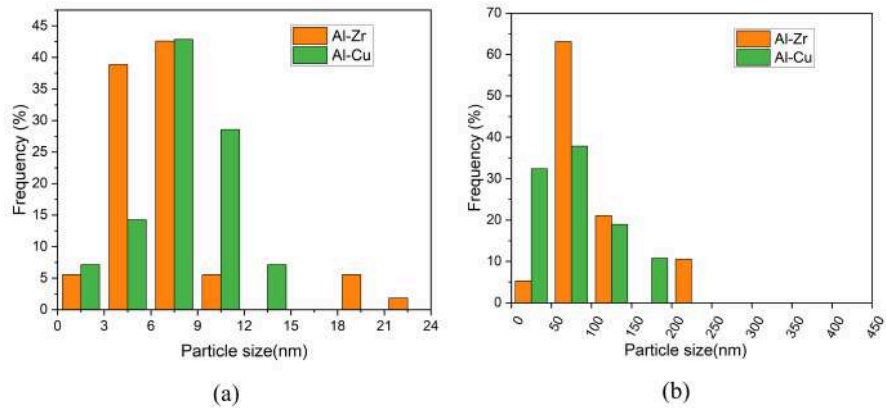

Figure 18. Precipitate size distribution in the as-built SLM processed Al 7075+ZrH$_2$ alloy. (a) Nanoprecipitates present in the grain interiors (d< 24 nm); (b) all other particles.

In summary, the addition of ZrH$_2$ microparticles resulted in substantial grain refinement, in further texture randomization, as well as in the formation of an abnormally high fraction of twin boundaries and in significant changes in the precipitation products. Altogether, these factors gave rise to a 30% increase in the hardness. The average Vickers hardness of the SLM processed Al 7075+ZrH$_2$ alloy amounted to 132.8±2.6 HV.

## 4. Discussion

*4.1 Icosahedral quasicrystal-enhanced nucleation*

The concept of crack reduction in selective laser melted non-weldable Al alloys by the addition of microparticles was introduced earlier by Martin et al. [33], who coated pre-alloyed gas-atomized Al 7075 and Al 6061 spherical powders with an average particle size of 45 μm with 1 vol.% hydrogen-stabilized zirconium nucleants using an electrostatic assembly technique. In that work, an average grain size of about 5 μm was reported for the as-built microstructures. A recent study by Tan et al. [45]

succeeded to engineer homogeneous, equiaxed Al 2024 microstructures, with an average size of 2 μm, by mixing the alloy powders with 0.7 wt.% of high purity Ti microparticles. Opprecht et al. [32] additionally proved the possibility to obtain submicrometer, crack free, microstructures in an Al 6061 alloy mixed with Yttrium Stabilized Zirconia (YSZ) nanoparticles. SLM of mixtures containing 2 vol. % of YSZ powder resulted nevertheless in fine bimodal microstructures, formed by columnar grains oriented with a <100> direction parallel to BD, and randomly oriented fine equiaxed grains, with diameters smaller than 1 μm. The current work further corroborates that a fully homogeneous and ultrafine-grained microstructure can be simultaneously achieved during SLM at a broad processing window by optimizing the fraction of $ZrH_2$ microparticles added to Al 7075 alloy powders. We also show that these microstructures exhibit minimal hot cracking, as well as a random texture, which is well known to favor isotropic mechanical behavior.

More importantly, our results evidence, additionally, that the grain boundary network of the SLM processed fine-grained Al 7075 + $ZrH_2$ samples contains an abnormally high number of twin boundaries (Figs. 10 and 13) in comparison with that corresponding to other polycrystalline cubic materials with random textures, which are usually well described by the classical MacKenzie distribution, and in which the fraction of such boundaries is smaller than 1 % [34]. To our knowledge, a similar excess fraction of twin boundaries has only been observed earlier following solidification of three metallic alloys with minute alloying additions that resulted, as in the present study, in significant grain refinement [46-48]. In particular, Kurtuldu et al. [46] reported a fraction of twin or near-twin grain boundaries exceeding 2 % in an Al–20 wt.%Zn alloy with small additions of Cr (1000 ppm) which was annealed under nearly isothermal growth conditions and then subjected to furnace cooling. In addition to increasing the fraction of twin boundaries, the addition of Cr also led to a drastic decrease in the size of the fcc grains. Similar observations were reported by Chen et al. [47] in commercial purity Al with Ti additions following annealing at 720 °C for 20 min and furnace cooling. Finally, iridium additions were observed to simultaneously reduce the grain size and promote the formation of multiple twinned grains in Au–28.4 at.%Cu–16.7 at.%Ag (yellow gold) polycrystals following casting at room temperature into Cu molds [48]. The fraction of twin boundaries produced was reported to increase up to 11 % when 200 ppm of Ir were added to the melt.

The origin of the abnormally high twin boundary fraction was attributed in the above mentioned studies [46-48] to the prevalence of a rarely observed solidification mechanism. The latter consists of the formation of icosahedral short-range order (ISRO) in the undercooled liquid, leading to the growth of icosahedral quasicrystals (iQC) which, in turn, act as a template for the nucleation of the matrix fcc grains [49]. This solidification mechanism, which is schematically shown in Fig. 18 [49], leads to an increase in the fraction of twin boundaries due to the coherent orientation relationships (ORs) developed between the icosahedral quasicrystalline particles and the fcc grains [50-52]. In particular, $\{111\}_{fcc}$ planes reportedly grow parallel to the triangular facets of the icosahedron. Minute alloying additions of certain elements such as Cr [46] and Ti [47] in Al or Ir [48] in yellow gold, would promote the formation of ISRO and thus would give rise to the observed large fraction of twin boundaries.

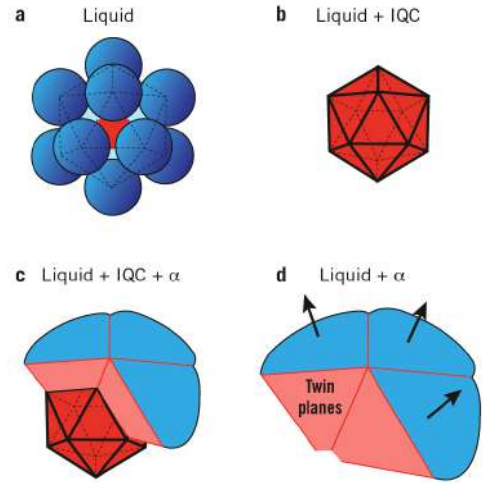

Fig. 19. Schematic illustrating the mechanism of icosahedral quasicrystal-enhanced nucleation.

In order to verify whether icosahedral quasicrystal-enhanced nucleation takes place during selective laser melting of the Al 7075+ $ZrH_2$ mixtures under investigation, we have analyzed individual neighboring orientations in as-built samples in the search for grain clusters with five-fold symmetry, emulating the methodology utilized for the same purpose in [46,48]. Fig. 20 illustrates the crystallographic relationships of 5 neighboring grains (depicted in Fig. 20(a)) found within the microstructure of the as-built Al 7075 + $ZrH_2$ sample. The {100}, {110}, and {111} pole figures of these five grains have been superimposed in Fig. 20(b), where it can be seen that they share a common <110> direction (highlighted using a red circle), which constitutes a five-fold symmetry axis. Moreover, multiple twin relationships have been found between pairs of grains within the analyzed cluster, as shown in Fig. 20(c). The corresponding twin axes have been highlighted using a white circle. Fig. 20 constitutes strong evidence that the five fcc grains, labeled 1–5, must have formed on a template having the 5-fold symmetry of a pentagonal pyramid. Based on the above crystallographic relationships, the facets of a regular icosahedron are numbered in Fig. 20 as the fcc grains that would have nucleated on them.

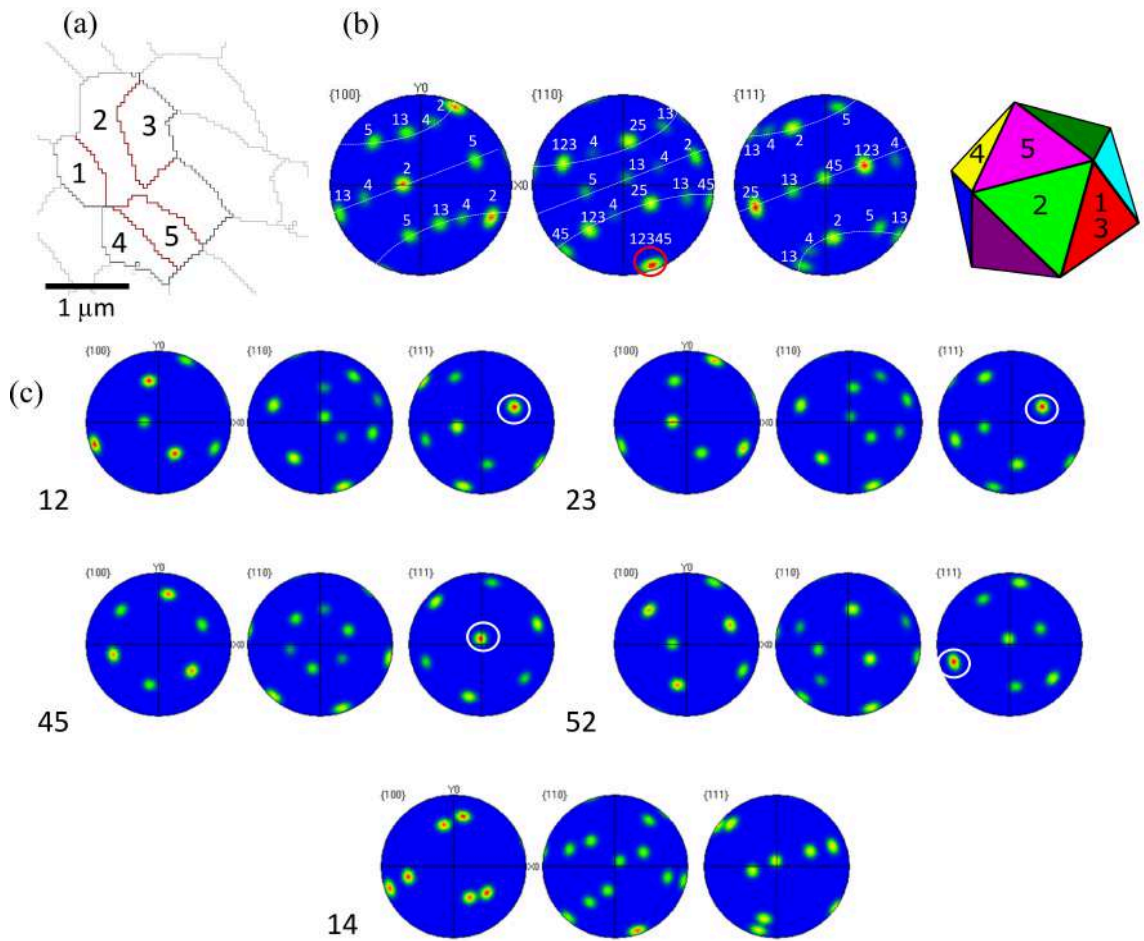

Fig. 20 (a) EBSD grain boundary map illustrating a cluster of five grains separated by twin boundaries (in red) and regular high angle boundaries (in black) belonging to the as-built Al7075 + ZrH2 sample; (b) global pole figures illustrating the five-fold symmetry around a <110> direction; (c) pole figures corresponding to several pairs of grains, in which 4 twin relationships (12,23,45,52) can be noticed. The <111> twin rotation axis has been highlighted using a white circle. A regular icosahedron with facets from which the fcc phase formed having the number as the grains in (a) is included.

Since quasicrystals were discovered by Schechtman et al. [53], who won the Nobel Prize for this work, they have been observed to coexist with crystalline particles in several binary and ternary Al alloys [54]. Quasicrystalline particles have, indeed, been recently observed in the microstructure of SLM processed Al 7075 powders [55]. An attempt was made to detect quasicrystalline particles by TEM in the as-built Al7075 + ZrH2 samples, with no success. The absence of quasicrystals in the fully solidified microstructure is, however, not surprising since, as described in detail in [48], the icosahedral quasicrystals forming in the supercooled liquid, and acting as a template for the growth of the solidified grains, can be the phases that are closest in free energy to the liquid, not the thermodynamically most stable ones for a given alloy composition [56-58]. For example, Kelton et al. [59] demonstrated that in a supercooled $Ti_{39.5}Zr_{39.5}Ni_{21}$ liquid the icosahedral short-range order first grows into a metastable iQC phase, then transforms into a more stable crystalline phase. The fact that earlier studies on icosahedral quasicrystal induced solidification in Al alloys have not reported the presence of iQCs suggests that these phases might only be stable for selected Al compositions. This would be in agreement with the current findings.

To our knowledge, the occurrence of icosahedral quasicrystal-enhanced solidification, and the associated abnormally high fraction of twin boundaries, have not been reported earlier in metallic alloys processed by SLM. It has been widely accepted that grain refinement by Zr alloying or by the addition of $ZrH_2$ particles to Al alloys occurs by the nucleation of submicron $L1_2$ $Al_3Zr$ phases, which act as inoculants for the Al matrix grains [33, 23-25]. Our research, however, evidences that, under the rapid solidification conditions inherent to SLM, quasicrystalline particles also contribute to the development of fine-grained microstructures by acting as effective inoculants in Al alloys.

*4.2 Outlook*

This research confirms that laser based manufacturing methods open new avenues for the investigation of fundamental metallurgical phenomena related to solidification and microstructure development. In particular, the influence of alloy composition on the occurrence of ISRO in the undercooled liquid and on the development of the grain boundary network are unexplored issues that might be facilitated by the use of additive manufacturing methods. Furthermore, the effect of SLM-specific grain boundary networks and of the presence of quasicrystalline particles on the properties of the fabricated alloys also remains unknown. Altogether, this work further confirms the added versatility provided by additive manufacturing methods for the design of advanced metals.

**5. Conclusions**

The aim of this work is to investigate solidification and microstructure development during SLM of Al 7075 powders, with and without $ZrH_2$ microparticle additions. Optimization of the volume fraction of microparticles and of the SLM parameters is followed by a thorough microstructure characterization at several length scales. This work has allowed unveiling the occurrence of unsuspected solidification mechanisms during the non-equilibrium conditions inherent to the SLM process. The following conclusions can be drawn from this study:

1. A fully isotropic microstructure, free of hot cracking defects, can be obtained by the addition of appropriate amounts of $ZrH_2$ microparticles to powders of the unweldable Al 7075 alloy and, simultaneously, by careful optimization of SLM process parameters. In particular, a homogeneous, equiaxed, ultrafine grained microstructure with a random texture is achieved by processing a mixture containing a 2.5 % vol. fraction of microparticles using a power value of 321 W, a hatch distance of 80 μm, and a scan speed of 1000 mm·$s^{-1}$.

2. The addition of microparticles leads to the development of distinct microstructural characteristics on the SLM processed samples, including significant grain size refinement, down to the ultrafine regime, an abnormally high fraction of twin boundaries that largely exceeds that corresponding to a random distribution of cubes, and the presence of grain clusters with five-fold orientation symmetry. All these microstructural features are absent in the SLM processed ==microparticle==-free microstructures.

3. Altogether, the above observations evidence that solidification during selective laser melting is enhanced by icosahedral quasicrystal-based nucleation. This mechanism has only been proven experimentally in conventionally cast Al-Zn and yellow gold alloys with minute additions of Cr (Kurtuldu et al., 2013) or Ti (Chen et al. 2018), and Ir (Kurtuldu et al., 2014), respectively. To our knowledge, this mechanism has not been reported before in SLM processed alloys.

**Acknowledgments**

MTPP wants to acknowledge the MAT4.0-CM project funded by Madrid region under program S2018/NMT-4381.